# Patterning of Fine Features on Material Surfaces Using a Ga Ion-Beam in a FIB-SEM

Supriya Ghosh[1*] and K. Andre Mkhoyan[1*]

[1]*Department of Chemical Engineering and Materials Science, University of Minnesota, Minneapolis, MN 55455, USA.*

**Abstract**

Since an ion-beam is a viable alternative to other material surface patterning techniques, a study of the structure, composition and dimension of patterned lines patterned on surfaces of Si and SrTiO$_3$ wafers with a Ga ion-beam was carried out. A combination of top-view SEM and cross-sectional STEM imaging and EDX spectroscopy applied to the patterned lines showed that the total ion-dose is the key parameter, which can be controlled by the degree of overlap between adjacent spots, beam dwell time at each spot, and the number of beam-passes for given beam size and current. The dependence between the used ion-doses and the patterned lines sizes was quantified. At higher ion-doses (> 10$^{15}$ ions/cm$^2$), the Ga ions remove part of the material in the exposed area creating 'channels' surrounded with amorphized regions whereas, at lower ion-doses only amorphization occurs, creating 'ridges' on the wafer surface. To pattern lines with similar sizes, an order of magnitude different ion-doses were required for Si and SrTiO$_3$ indicating strong material dependence. Lines as fine as 10 nm can be reproducibly patterned and characterized on the surfaces of materials, when low ion-doses are used, typically in the range of 10$^{14}$-10$^{15}$ ions/cms$^2$.

**Keywords:** Patterning, ion-beam lithography, FIB, line width, SEM, STEM

**\*Corresponding Authors.** E-mail: ghosh115@umn.edu, mkhoyan@umn.edu

**Introduction**

Nanoscale patterning on material surfaces has a huge demand in a wide range of applications ranging from semiconductor device fabrications to creation of micro-channels in nano-fluidics studies (*Fang et al., 2017; Liu et al., 2019*). Ion-beam lithography is one such technique, allowing fabrication of patterns with fine features at the nm-length scales owing to versatility in the choice of the available ion sources as well as the beam energies that can be employed during patterning (*Bassim et al., 2014; Gierak et al., 2009; Kim et al., 2012; Langford et al., 2007; Li et al., 2021; Li et al., 2019; Ocola et al., 2014; Petit et al., 2005; Tseng, 2004; Tseng, 2005*). It has been successfully used for mask repairing and circuit editing in semiconductor device fabrication and testing (*Li et al., 2021*), creation of periodic surface arrays for nanophononics (*Li & Liu, 2018; Manoccio et al., 2021; Seniutinas et al., 2016*), formation of nucleation sites needed for nanoparticle growth (*Gierak et al., 2009; Gierak et al., 2005*), and patterning magnetic orders in thin films (*Suharyadi et al., 2016; Urbánek et al., 2018*). Ion-beam lithography provides comparable resolutions to that available in optical or electron-beam lithography systems without requiring the application of resists or masks (*Cuerno & Kim, 2020; Gierak et al., 2005; Petit et al., 2005; Sabouri et al., 2015*). One instrument that can be used for this is a focused ion beam (FIB) system, which is now used to study nanomaterials including imaging using the ion beam, removing and depositing material (*Moll, 2018; Volkert & Minor, 2007*), and preparing ultra-thin lamellae samples for transmission electron microscopy (TEM) (*Giannuzzi & Stevie, 1999; Mayer et al., 2007; Zhong et al., 2021*).

When high energy ions interact with the material surface, the ions can transfer their energy to the atoms on the surface causing them to sputter (*Giannuzzi et al., 2005*). This enables the ion-beam to be utilized as a tool for milling. Since ions-beams can be produced in FIB systems with very high current densities with nanometer dimensions and have relatively smaller penetration depth into the material, as compared to electron beams (*Kim et al., 2012*), precise removal of material can be achieved. The efficiency of material removal, which depends on the sputtering yield of the material, can be related to the ion-flux and adjusted in the FIB. Hence, this technique can be used to pattern different types of materials including harder materials like semiconductor ceramic oxides (*Tseng, 2004*). However, it is critical to evaluate and understand the effects of ion-beam on the structural and compositional changes in the exposed areas of the materials, as they set the limits of the achievable patterning resolution.

In this work, a study of the impact of key ion-beam parameters on the structure and resolution of patterned lines was carried out with the goal to achieve the highest resolution or thinnest patterned line. We investigated the effects of beam dwell time at each position, the degree of overlap between adjacent ion-beam positions, and the number of ion-beam passes used to create the patterned line. Two different materials, Si and SrTiO₃ were considered here to assess material-specific effects of the Ga ion-beam. Analysis of the structure, composition and sizes of patterned lines were carried out using scanning electron microscopy (SEM) for top-view imaging and high-angle annular dark-field scanning transmission electron microscopy (HAADF-STEM) for cross-sectional imaging and energy dispersive X-ray (EDX) spectroscopy for elemental mapping.

**Materials and Methods**

The Ga ion-beam patterning was carried on a FEI Helios G4 dual-beam FIB-SEM system. The FIB-SEM chamber was plasma cleaned for 2 hours prior to each experiment, to remove any potential contaminants in the chamber. The wafers of (110) Si and (001) SrTiO₃ used for this study were cleaned using solvents (isopropyl alcohol, acetone, and distilled water) and air dried prior to loading the into the FIB-SEM. To reduce carbon contamination, the wafers were directly mounted onto screw-top SEM stubs (from Ted Pella Inc.) without carbon tape.

All the patterned lines discussed here were created using a $E_0 = 30\ keV$ Ga ion-beam, which is the standard beam energy utilized for milling operations. To achieve the finest possible ion-beam probe, the smallest beam forming aperture was selected, which produces an ion-beam of diameter 5.3 nm and current $I_0 \simeq 1\ pA$ (measured directly using a Faraday cup). To avoid beam broadening, the ion-beam incidence angle to the wafer surface was set to be 90°, with the wafer surface being the focal plane. The ion-beam parameters and their considered ranges were: (1) the degree of overlap between the ion probes in neighboring pixels or "beam overlap" from -200 to 95 %; (2) beam dwell time at each pixel, from 25 ns to 1 µs; and (3) the number of beam scans per pattern line or "beam passes", from $10^2$ to $10^6$. Pattern lines 10 µm in length with a separation distance between lines ranging from 0.5 to 2 µm were made on both Si and SrTiO₃ wafers. It should be noted that the wafer regions used for patterning had no prior exposure to the Ga ion-beam. The patterning was monitored by SEM using the secondary electron (SE) detector for imaging.

To study the impact of the Ga ion-beam on both Si and $SrTiO_3$ wafer surfaces, high-magnification SEM images from each set of patterned lines were obtained. These SEM images were recorded using the back-scattered electron (BSE) detector, operated in the immersion mode, for high contrast and better visibility of the in-plane surface modifications. An incident electron beam with energy of 15 keV, probe current of 0.10 nA, and working distance of 3 mm was used.

To study the profiles and the atomic structure of the patterned lines, cross-sectional samples were prepared using the standard lift-out method in the FIB and evaluated using STEM. Before FIB sectioning, a 50-nm-thick protective layer of Pt was deposited in the region of interest followed by deposition of 2 µm of protective amorphous carbon (a-C) layer. The final thinning of these cross-sectional FIB-cutout STEM samples was performed using a 30 keV Ga ion beam followed by a 2 keV ion-beam to remove any surface damaged layers. HAADF-STEM images of the patterned lines were recorded using an aberration corrected FEI Titan G2 (S)TEM 60-300 microscope. The STEM is equipped with a CEOS-DCOR probe corrector, monochromator and Super-X EDX spectrometer. The microscope was operated at 200 keV, with a probe convergence angle of 25.5 mrad with a HAADF inner and outer collection angles of 55 and 200 mrad respectively. A probe current of 120 pA was used for both imaging and EDX compositional maps.

**Results and Discussion**

Schematic description of the set-up used for patterning in the dual-beam FIB-SEM system and description of the ion-beam parameters are presented in Figure 1 (a, b). The characteristics of the patterned lines are sensitive to the ion-dose delivered ($D_I$), as it controls the effective volume of material being removed. While the ion-beam current (selected by apertures in the system) controls the size of the probe, the other parameters (beam dwell time, beam overlap, and number of beam passes) determine the effective ion-dose delivered to the pattern lines. For example, shorter beam dwell times would need larger number of beam-passes over a pattern line to deliver the same ion-dose. It should be noted that dual-beam FIB-SEM systems also have an in-built pattern milling parameter, often referred as the "z-parameter", for controlling the ion-doses. The relations between these parameters are described in Figures 1 (c-e). In such dual-beam FIB-SEM systems, the desired patterns can be applied to the wafer surface either using the in-built "pattern/array" function or from imported pattern files with encoded ion-beam parameters. Figure 1(f) shows examples of different patterns created on the surface of a (110) Si wafer using different ion-doses. As can be seen from these SEM images, uniform patterned lines with varying sizes and dimensions can be made reliably and reproducibly.

To evaluate the structure of the patterned line, top-view SEM images from the lines created at different ion doses and high-resolution cross-sectional STEM images from the same lines were obtained and analyzed. Figure 2 shows two pairs of such SEM-STEM images from lines patterned on (001) SrTiO$_3$ wafer with a relatively high ion-dose of $D_I = 4\times10^{18}$ ions/cm$^2$ (or 'z'=1000), and on (110) Si wafer with a low ion-dose of $D_I = 2.2\times10^{16}$ ions/cm$^2$ (or 'z'=5). At higher Ga ion-doses, some of the wafer material is removed, which is referred in the following discussions as formation of a 'channel'. The impact of Ga ions also results in amorphization of the crystal in the surrounding regions. These amorphous regions are directly observable in the HAADF-STEM images, as they appear darker due to (i) slightly lower density of material, and (ii) lack of crystallinity for strong electron-beam channeling, as can be seen in Figures 2 (b). At lower ion-doses, when the volume of material removed is much smaller (or none), only amorphization of the impacted area is observed (Figure 2 (e)). The interface, as expected, appears partially crystalline and partially amorphous. STEM-EDX elemental maps corroborate with these observations (Figure 2 (c) and (f)). These elemental maps also suggest that some Ga ion implantation should be expected in the amorphous regions of the patterned lines, where the degree of Ga ion accumulation is a function of ion-dose and wafer material (*Nastasi et al., 1996*).

To quantify the dimensions of these patterned lines, the following parameters are introduced (Figure 2): $w_1$ - the width of the channel from SEM images (darker area); $w_{1,t}$ - the total width of the ion-beam impacted area from SEM images, which includes the channel and amorphized surroundings; $w_{2,s}$ - the width of the channel at the wafer surface level from STEM images; $w_{2,b}$ - the width at the bottom of the channel from STEM images; $w_{2,t}$ - the total width of the ion-beam impacted area from STEM images, which includes the channel and amorphized surroundings; $d_t$ - the total depth of the 'channel' from STEM images; $h_r$ - the height of the 'ridge' from STEM images. These parameters are used to evaluate the resolution of patterned lines for different ion-beams.

To evaluate the widths of the pattern lines as a function of ion-dose ($D_I$), first a (110) Si wafer was used where a set of lines were made using: (i) different number of beams passes, and (ii) different degrees of beam overlap applied to each line. In both sets, the same dwell time of 25 ns/pixel was used. The results are presented in Figure 3, where SEM images of the lines made with same ion-doses, achieved either by adjusting the number of beams passes or by the degree of beam overlap, look very similar (Figure 3 (b)). The actual line widths obtained from these SEM images are presented in Figure 3(d), where it can be seen that both the width of the patterned line ($w_1$) and the total ion impacted region ($w_{1,t}$) depend on the ion-dose. While the width of the patterned line is much more sensitive to ion-doses used when compared to total impacted region, both widths are independent of number of passes or the degrees of beam overlap used. When the widths of the same patterned lines were evaluated more precisely from cross-sectional HAADF-STEM images (shown in Figure 3 (c)), it showed consistency among values for the total width of the ion-beam impacted regions obtained from STEM and SEM images (Figure 3 (e)).

For line created at higher ion-doses ($D_I > 10^{16}$ ions/cm$^2$), the cross-sectional HAADF-STEM images provide two values for the width of the patterned lines, $w_{2,s}$ and $w_{2,b}$, due to the asymmetric profile of the channel with depth. Thus, no direct quantitative comparison can be made with the values obtained from SEM images. However, $w_1$ from SEM images is in-between values of $w_{2,s}$ and $w_{2,s}$ (Figure 3 (e)). The analysis of the depth of each channel, obtained from cross-sectional HAADF-STEM images, shows strong dependence of $d_t$ on the ion-beam doses (Figure 3 (f)). It should be noted that at lower ion-doses, below $2.2 \times 10^{16}$ ions/cms$^2$, only amorphization of Si takes place, which results in a formation of a "ridge" on the surface due to differences in densities ($\rho_a^{Si} = 2.28$ g/cm$^3$ vs $\rho_c^{Si} = 2.33$ g/cm$^3$), $\rho_a^{Si} = 0.97 \rho_c^{Si}$, resulting in slight volume expansion over the surface.

Next, the effect of the ion-beam dwell time on the widths of patterned lines on Si was evaluated. Three beam dwell times of 25, 500 and 1000 ns were considered using a 50 % beam overlap. SEM images and the dependence of the line width on dwell time are presented in Figure 4. Interestingly, while the lines made with 500 and 1000 ns dwell times are very similar when the same ion-dose is used, the lines made with 25 ns dwell time are considerably narrower. While the origin of this line width dependence on dwell time is not clear, we speculate that the characteristic time for bond breaking and/or crystal-to-amorphous transformation in Si is comparable with 25 ns resulting in dramatic reduction of Ga ion interaction volume at lower dwell times as well as scope of more secondary scattering events at higher dwell times, increasing interaction volumes. Further in-depth evaluation is needed for proper understanding of this observation. The finest identifiable line that was patterned on Si with this method is ~ 10 nm, using a Ga ion beam dose of $D_l$ =5×10$^{14}$ ions/cm$^2$ and dwell time of 25 ns.

Further, to evaluate differences in pattern line resolutions that can be attained in different materials, an additional study was carried out using a perovskite SrTiO$_3$ wafer. Several sets of lines were patterned on the (001) SrTiO$_3$ wafer using the same ion-beam parameters as used for Si wafer previously. Complimentary SEM and HAADF-STEM images from the set of lines made on SrTiO$_3$ wafer with different ion-doses are shown in Figure 5 (a) and (b). These results can be directly compared with the results obtained for Si wafer (Figure 3 (b and c)). While the overall structure of the lines in SrTiO$_3$ and Si are similar, the variations in contrast in both sets of (SEM and STEM) images are different. As in case with Si, the regions with darker contrast in cross-sectional HAADF-STEM images correspond to amorphous SrTiO$_3$. The noticeable differences between patterned lines in SrTiO$_3$ and Si, as can be seen in HAADF-STEM images, are in the formation of amorphous regions which produces different contrast variations in the top-view SEM images (Figure 3 (b) and 5 (a)). In SrTiO$_3$, 'channels' are formed above ion-doses of $D_l$ =4.3×10$^{17}$ ions/cm$^2$, which is an order of magnitude higher than that seen in Si ($D_l$ =2.2×10$^{16}$ ions/cm$^2$). Additionally, since the density of amorphous SrTiO$_3$ is considerably lower from crystalline SrTiO$_3$ ($\rho_a^{STO}$ = 3.84 g/cm$^3$ vs $\rho_c^{STO}$ = 5.11 g/cm$^3$), $\rho_a^{STO}$ = 0.75$\rho_c^{STO}$ (*Medvedeva et al., 2022*), the 'ridges' forming here at lower ion-doses are considerable higher than dose observed in Si.

The quantification of the patterned line structures made on SrTiO$_3$ wafer is presented in Figure 5 (c-e), where the dependence of line width with the ion-dose has similar trends to that seen in Si. Similar behavior is also observed for the 'channel' depth (d$_t$) and 'ridge' height (h$_r$) as a function of ion dose, with a different turning point (or characteristic ion-dose) required to transition from channel formation to ridges on the surface (Figure 5(d)). Widths of lines created on SrTiO$_3$ wafer at different dwell times of 25, 100, and 500 ns resulted in very close width versus dose behavior suggesting that, unlike Si, characteristic time for crystal-to-amorphous transformation in SrTiO$_3$ should be much shorter than 25 ns. The quantitative comparisons of the line widths and 'channel' depths between SrTiO$_3$ and Si is presented in Figure 6. The actual widths of the lines patterned on SrTiO₃ wafer, determined from more accurate HAADF-STEM images, are about x2 larger, while the depths of the 'channels' in SrTiO₃ is about x2 smaller than in Si. We speculate that this is due to three parameters: average atomic number in the materials ($Z_{(STO)} = 17$ and $Z_{Si} = 14$), average interatomic bonding energy ($E_{(STO)} = 4.00$ eV and $E_{Si} = 2.32$ eV) (*Eglitis & Vanderbilt, 2008*), and atomic density of material ($\rho_{STO} = 5.11$ g/cm³ and $\rho_{Si} = 2.33$ g/cm³). Combination of these three parameters yields considerably different Ga ion penetration depth and sputtering rates of atoms from material for SrTiO₃ and Si. The finest identifiable line that was patterned in SrTiO₃ with this method is about 10 nm, when Ga ion beam dose of $D_I = 1.2 \times 10^{14}$ ions/cm². While even finer, < 10 nm, lines can be patterned with lower ion-doses, their identification on the surface is likely to be the limiting factor.

**Conclusion**

In this study using a Ga ion beam in a dual-beam FIB-SEM system, sets of lines were patterned on the surfaces of Si and SrTiO₃ wafers. Top-view SEM and cross-sectional STEM images obtained from the same patterned lines showed that the dominating parameter is the total dose of the ion-beam, and it can be adjusted using: (i) beam current and aperture size (as a starting point), (ii) the degree of overlap between adjacent spots, (iii) beam dwell time at each spot, and (iv) the number of beam passes. It was observed that at high ion-doses, the Ga ions remove part of the material in the exposed area creating lines that resemble 'channels' surrounded with regions of amorphized crystal, and at low ion-doses only amorphous lines are made with 'ridge'-like appearance on the surface. Detailed quantification of the dimensions of patterned lines from SEM and HAADF-STEM images showed that there is strong dependence between ion-doses used and size of the lines made, and it can vary orders of magnitudes depending on the material: patterning the same size line in SrTiO₃ takes x10 more ion-dose than in Si. The quantification also showed that fine lines as small as 10 nm wide can be reproducibly patterned and characterized on the surfaces of SrTiO₃ and Si, and of any other material, when ion-beam diameter was about 5 nm. For such fine lines low ion-doses are needed, typically in the range of $10^{14}$-$10^{15}$ ions/cm² at a beam energy of 30 keV. While even finer amorphous lines can be patterned, SEM imaging is the limiting factor for their identification and quantification.

**Figures:**

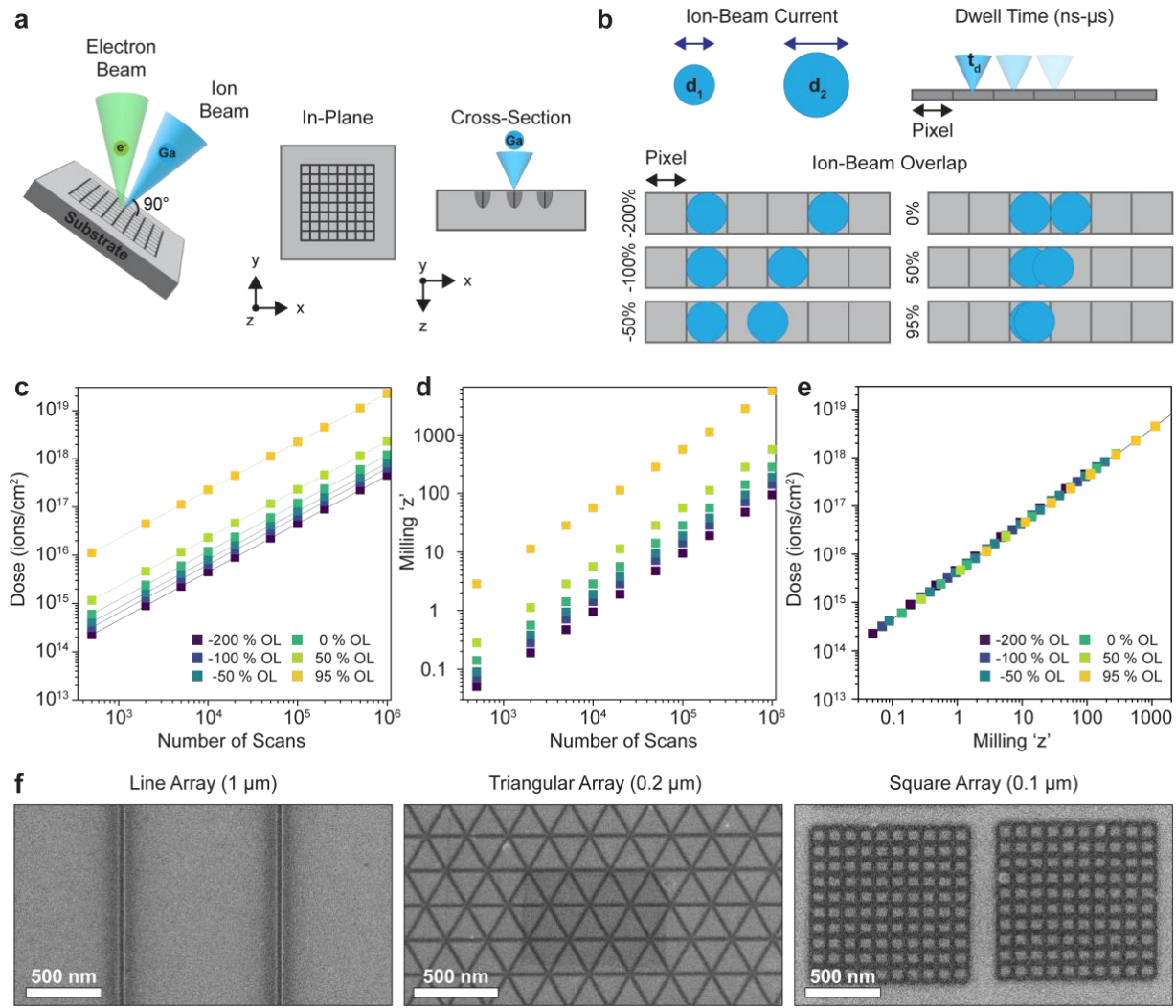

**Figure 1: Patterning in a dual-beam FIB-SEM.** (a) Schematics of the electron- and ion-beam orientations during patterning of the substrate using a dual-beam FIB along with illustration of the Ga beam impact on substrate in-plane and in cross-section. (b) Ion-beam parameters for pattern milling: beam current controlling the beam diameter, the dwell time and the beam overlap. (c-e) Relationships between ion-dose ($D_I$), the number of beam scans (or passes) across the pattern line, beam overlap, and 'z-parameter'. (f) Examples of patterns created on a (110) Si wafer surface using a 30 keV Ga ion beam, with 1 pA, 25 ns dwell time and a beam overlap of -200%. Line array patterned with ion-dose of $D_I = 4.3 \times 10^{17}$ ions/cm$^2$, and triangular and square arrays patterned with ion-dose of $D_I = 4.3 \times 10^{16}$ ions/cm$^2$.

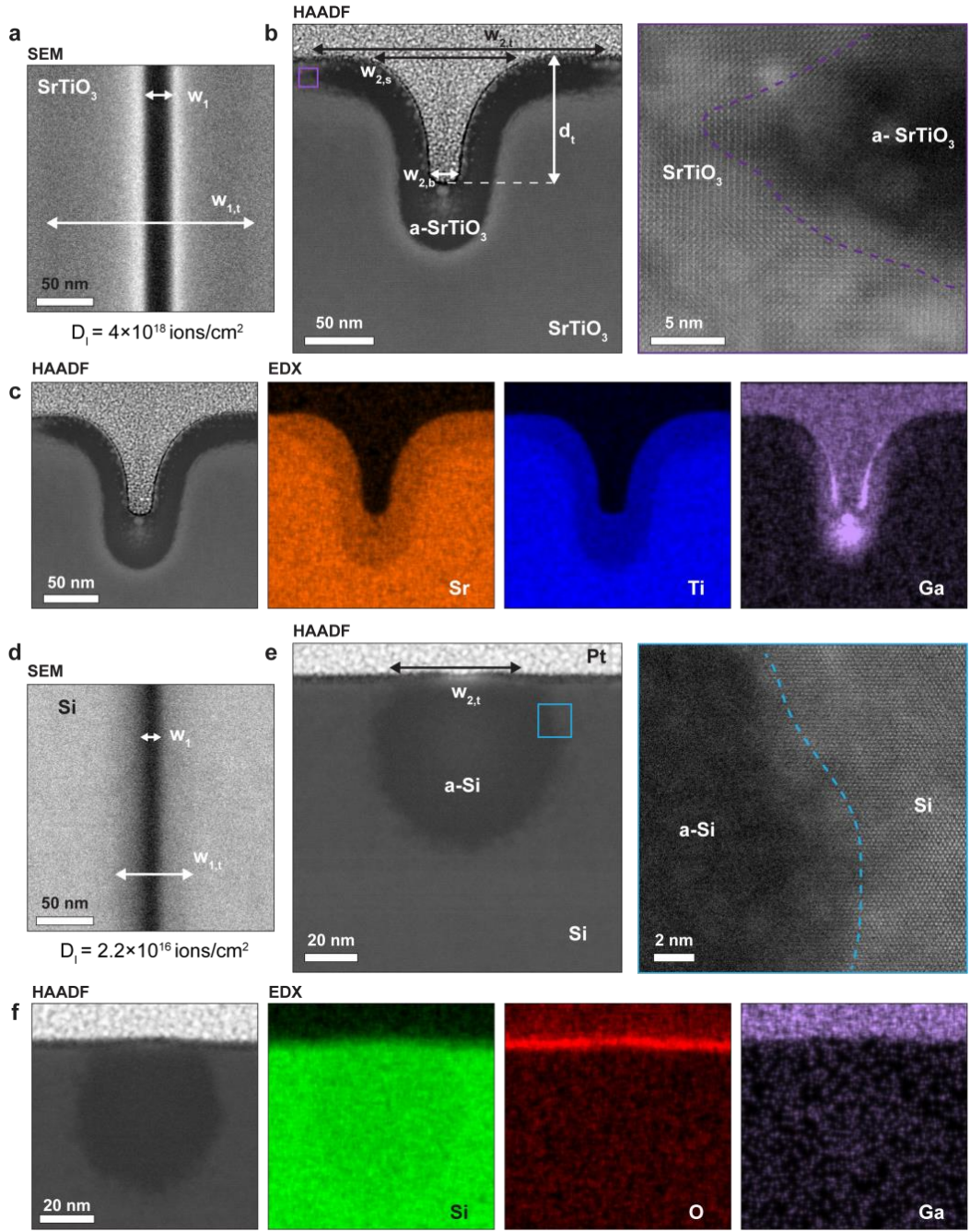

**Figure 2: Impact of ion-beam patterning on material structure.** (a) Top-view SEM image of patterned line made on (001) SrTiO₃ wafer surface with high ion-dose of $D_l = 4×10^{18}$ ions/cm². (b) Cross-sectional HAADF-STEM images of the patterned line in (a), where the regions with dark contrast correspond to amorphized SrTiO₃. (c) STEM-EDX maps from the patterned channel in SrTiO₃ showing distribution of Sr and Ti as well as incorporation of some Ga in the amorphous material. (d) Top-view SEM image of patterned line made on (110) Si wafer surface with low ion-

dose of $D_I = 2.2 \times 10^{16}$ ions/cm². (e) Cross-sectional HAADF-STEM images of the patterned line in (d), where the dark contrast of the line is due to amorphization of Si. (f) STEM-EDX maps from the patterned line in Si showing distribution of Si, surface O, and barely detectable Ga in the amorphous region. The parameters used to evaluate the patterned line resolution are described with arrows in (a, b).

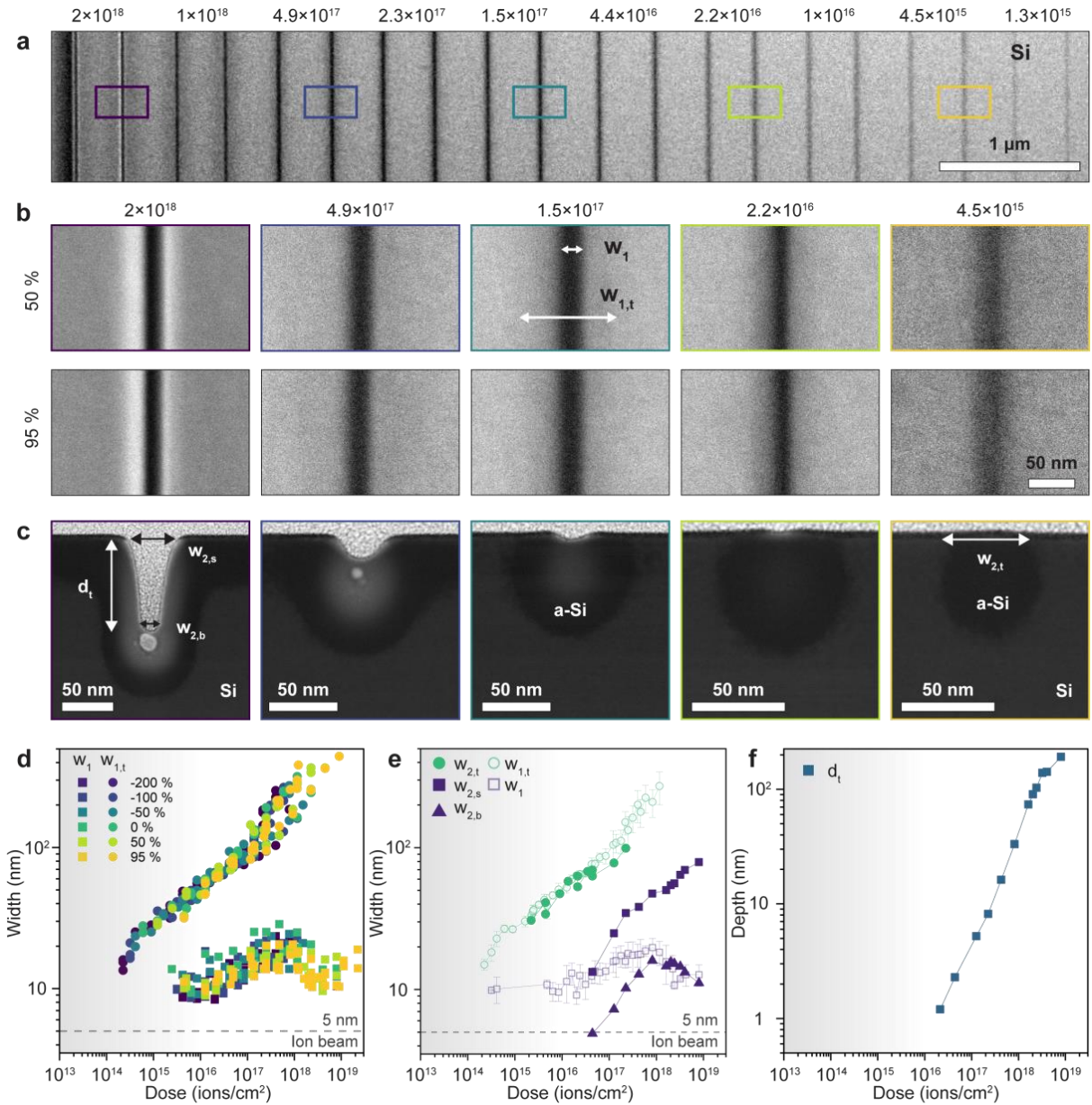

**Figure 3: Resolution of patterning in Si.** (a) SEM image of a set of patterned lines made on (110) Si wafer with different ion-doses. Here 50% beam overlap was used. The ion-doses are in units of [ions/cm²]. (b) SEM images comparing patterned lines with 50% and 95% beam overlaps at different $D_I$ as indicated. (c) Cross-sectional HAADF-STEM images of patterned lines made on Si at same $D_I$ as shown in (b). Here, the darker contrast around the outer edge of pattern area is due to amorphized Si. The lighter contrast around the center, visible for lines made with higher ion-doses, is due to implantation of heavier Ga ions. (d) Width of the line, $w_1$, and the ion-beam-impacted area, $w_{1,t}$, as a function of ion dose at different beam overlaps measured from SEM images. (e) Width of pattern lines measured from the HAADF-STEM cross-section images as a function of ion dose. For comparison the average widths measured from SEM images are also shown. (f) Depth of the 'channel' as a function of ion-doses. Region in gray in (d-f) represents the ion-doses at which mainly amorphization is observed.

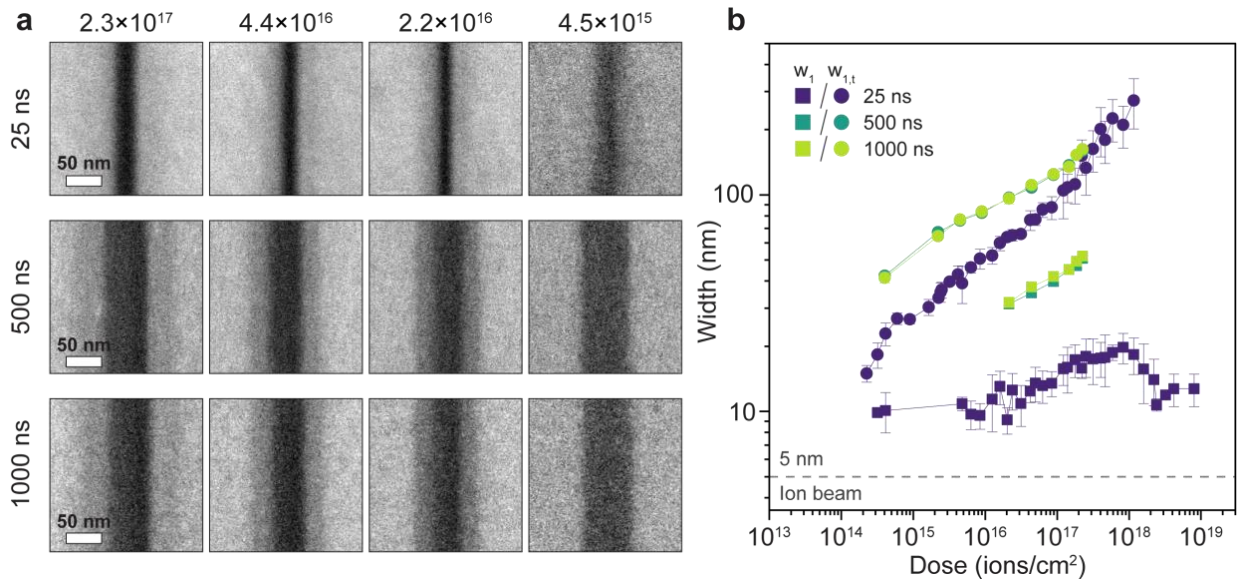

**Figure 4: Impact of dwell time on Si patterning.** (a) A set of SEM images of patterned lines made with different ion-doses and dwell times. Here 50% beam overlap was used. The ion-doses shown on the top are in units of [ions/cm$^2$]. (b) Width of patterned line as a function of ion-doses for beam dwell times of 25, 500 and 1000 ns measured from SEM images.

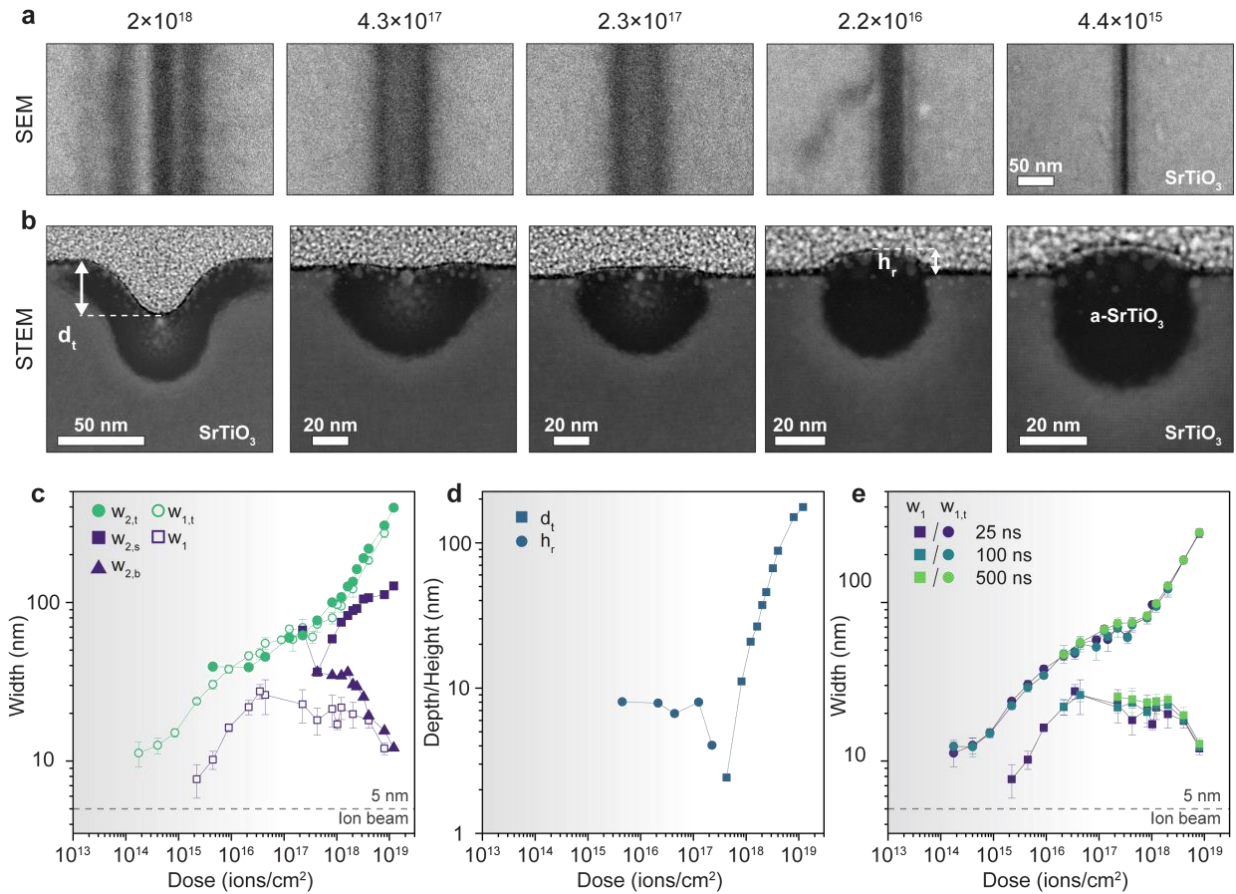

**Figure 5: SrTiO₃ patterning resolution.** (a) A set of top-view SEM images showing the patterned lines on the surface of (001) SrTiO₃ wafer made with different ion-doses. Here -200 % beam overlap and 25 ns dwell time were used. (b) Cross-sectional HAADF-STEM images of the lines. The darker contrast around the outer edge of patterned area is due to amorphized SrTiO₃. The areas with lighter contrast are due to implantation of heaver Ga ions. At high ion-doses a 'channel' forms surrounded by amorphous regions. At low ion-doses, only SrTiO₃ amorphization is seen accompanied with formation of a 'ridge' on the surface. (c) Width of the line measured from the top-view SEM and cross-sectional HAADF-STEM images as a function of ion-dose. (d) Depth of the 'channel' and height of the 'ridge' measured from cross-sectional HAADF-STEM images of the lines as a function of ion-dose. (e) Width of the line measured from the top-view SEM images as a function of ion-dose for three different dwell times used during patterning.

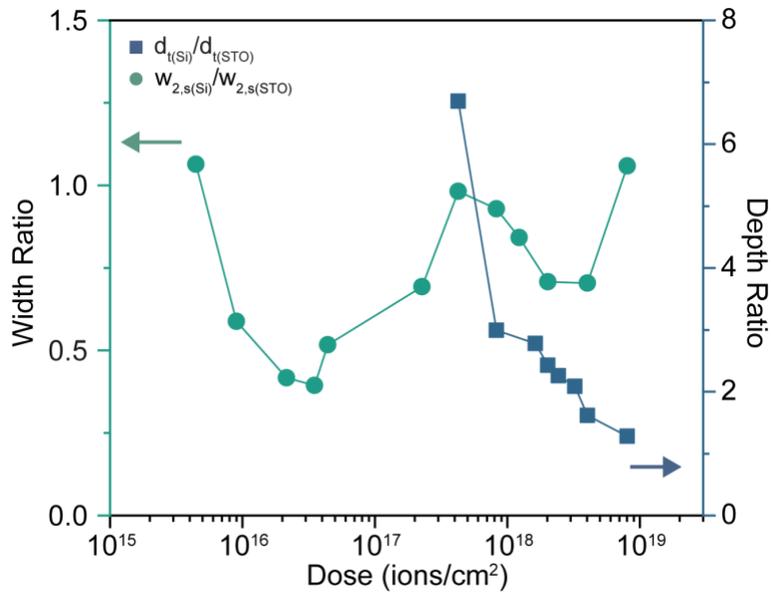

**Figure 6: Comparison of line widths and depths.** The ratio of widths and depths of pattern lines, measured from cross-sections HAADF-STEM images, between Si and SrTiO₃ wafers as a function of ion-dose. They are calculated from the data shown in Fig. 3(e, f) and Fig. 5(c, d).